\documentclass[twocolumn,aps,prl,groupaddress]{revtex4-1}
\usepackage{amsmath}
\usepackage{amssymb}
\usepackage{graphicx}
\usepackage{hyperref}

\usepackage[english]{babel}
\usepackage{times}

\makeatletter

\@ifundefined{textcolor}{}{%
 \definecolor{BLACK}{gray}{0}
 \definecolor{WHITE}{gray}{1}
 \definecolor{RED}{rgb}{1,0,0}
 \definecolor{GREEN}{rgb}{0,1,0}
 \definecolor{BLUE}{rgb}{0,0,1}
 \definecolor{CYAN}{cmyk}{1,0,0,0}
 \definecolor{MAGENTA}{cmyk}{0,1,0,0}
 \definecolor{YELLOW}{cmyk}{0,0,1,0}
}

\newcommand{\prlsection}[1]{{\em {#1}:--~}}

\DeclareRobustCommand{\openzero}{\leavevmode\hbox{0\kern-.55em0}}
\mathchardef\minus="002D

\newcommand{\1}{\leavevmode{\rm 1\ifmmode\mkern  -4.8mu\else\kern -.3em\fi I}}

\begin{document}

\title{Coherence generating power of quantum unitary maps and beyond}

\author{Paolo Zanardi, Georgios Styliaris, and Lorenzo Campos Venuti}

\affiliation{Department of Physics and Astronomy, and Center for Quantum Information
Science \& Technology, University of Southern California, Los Angeles,
CA 90089-0484}

\begin{abstract}
Given a preferred orthonormal basis $B$ in the Hilbert space of a
quantum system we define a measure of the coherence generating
power of a unitary operation  with respect to $B$. 
This measure is the average coherence generated by the operation acting 
on a uniform ensemble of incoherent states.
We give its explicit analytical form in  any dimension and provide an operational
protocol to directly detect it. 
We characterize the set of unitaries with maximal coherence generating power and study the properties
of our measure when the unitary is drawn at random from the
Haar distribution. For large state-space dimension a random unitary
has, with overwhelming probability, nearly maximal coherence generating
power with respect to any basis. Finally, extensions to general unital
quantum operations  and the relation to the concept of asymmetry
are discussed.
\end{abstract}

\maketitle

\prlsection{Introduction} One of the most fundamental attributes
of quantum dynamical systems is their ability to exist in linear superpositions
of different physical states. In fact any pure quantum state can be
regarded, in infinitely many different ways, as a linear superposition
of a {\em{basis}} of distinguishable quantum states. The experimental
signature of such a superposition structure (in the given basis) is
known as {\em{quantum coherence}} \cite{glauber_coherent_1963}. The
latter is also known as one the basic ingredients for quantum
information processing \cite{nielsen_quantum_2000} and its protection e.g.,
by decoherence-free subspaces \cite{zanardi_noiseless_1997,zanardi_dissipation_1998,lidar_decoherence-free_1998}, is one of the fundamental
challenges in the field.

Over the last few years we have witnessed a strong renewal of interest
in the {\em{quantitative}} theory of coherence \cite{baumgratz_quantifying_2014,streltsov_quantum_2016}.
This is partly practically motivated by the role that quantum coherence
plays in quantum metrological protocols (see e.g., discussion in \cite{marvian_how_2016})
and, on a more conceptual ground, by its relation to the general resource
theory of asymmetry \cite{marvian_asymmetry_2014,marvian_extending_2014,marvian_mashhad_symmetry_2012}.
Quantum coherence is also believed to play a role in some fundamental
biological process \cite{engel_evidence_2007,collini_coherently_2010,lambert_quantum_2013} as well as in quantum thermodynamics
\cite{lostaglio_quantum_2015,lostaglio_description_2015}. The general idea
is that one can quantify quantum coherence by introducing a real-valued
function over the quantum state-space, a {\em{coherence measure}},
such that it vanishes for all the states that are deemed to be {\em{incoherent}}
and cannot increase under some class of operations that preserve incoherence
\cite{levi_quantitative_2014}. Even if a preferred basis is chosen the choice
of the coherence measure is not unique and different options have
been discussed in the literature \cite{marvian_how_2016,baumgratz_quantifying_2014,girolami_observable_2014,yao_frobenius-norm-based_2016}.

In this paper we address a closely related problem, which was first tackled in \cite{mani_cohering_2015}: the quantification
of the power of a quantum operation to generate
coherence. Again, even when an underlying coherence measure is assumed,
the definition of the {\em{coherence generating power}} (CGP)
of a Completely Positive (CP)-map is not unique and different lines of attack are possible \cite{bromley_frozen_2015,mani_cohering_2015,bromley_frozen_2015,garcia-diaz_note_2015}
(see Sect IV C of \cite{streltsov_quantum_2016} for a comprehensive list of references).
All of these approaches, however, are cast in terms of an optimization problem
that is extremely hard to handle for generic channels in arbitrary dimensions.

\begin{figure}[t]
\begin{centering}
\includegraphics[width=0.9\columnwidth]{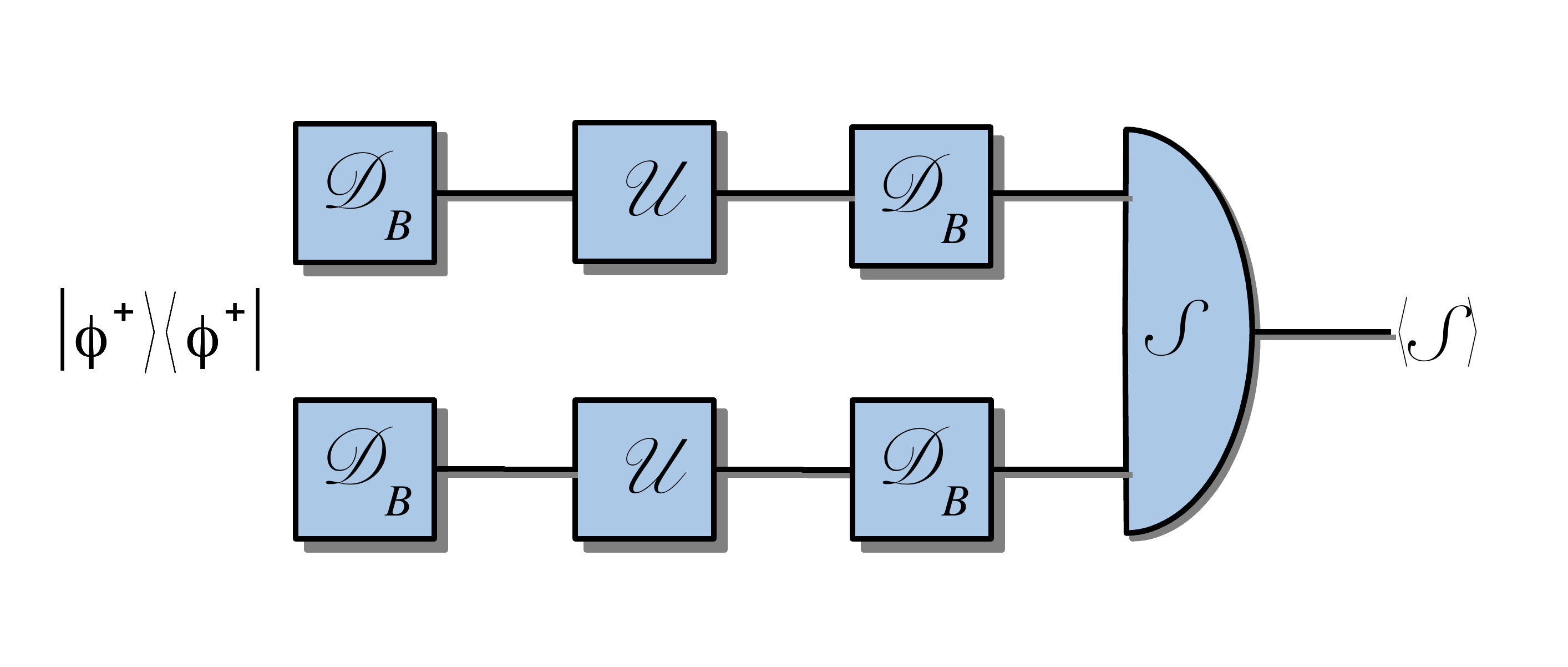}
\par\end{centering}
\vspace{-4mm}
\caption{Protocol for the direct detection of the Coherence Generating Power
(CGP) Eq.~\eqref{C_B} of the unitary CP map $\mathcal{U}$ based
on Eq.~\eqref{C_B-Texplicit}. Here $\mathcal{D}_{B}$ is the dephasing
super-operator for the preferred basis $B$, the measurement of the
swap operator is denoted by $\mathcal{S}$ and $|\Phi^{+}\rangle:=d^{-1/2}\sum_{i=1}^{d}|i\rangle^{\otimes2}$.
\label{protocol-Fig}}
\end{figure}

Following the spirit of Ref.~\cite{zanardi_entangling_2000} in entanglement theory, we shall here pursue
a different strategy based on probabilistic averages.
We define the CGP of a map as the average coherence that is generated
when the corresponding quantum operation is performed over a suitable
input ensemble of incoherent states. We shall here firstly focus on
unitary maps and introduce a  definition of CGP based on
a uniform ensemble (see below for a precise definition) of incoherent
states.

Our measure of CGP is analytically computable for arbitrary unitary map in any dimension. It also enjoys several natural
and desirable properties e.g., invariance under pre- and post-processing
by incoherent unitaries. We shall  present a simple operational protocol
for the direct detection of the CGP of a given map which does not
involve the ensemble generation or quantum process tomography \cite{chuang_prescription_1997,poyatos_complete_1997}.
The set of unitary operations with maximal CGP is easily characterized
and some universal statistical properties of our measure over the
group of unitaries can be established rigorously. We will also provide
some numerical study of the distribution of CGP in various dimensions
$d$ (for $d=2$ analytical form is available). Finally, extensions
of CGP to arbitrary unital operations are discussed as well as the connection
to the broader concept of asymmetry generating power of a map. The
proofs of the Propositions can be found in \cite{SM}.

\prlsection{Preliminaries} Let
$B=\{|i\rangle\}_{i=1}^{d}$ be an orthonormal basis in the Hilbert
space ${\cal H}\cong{\mathbf{C}}^{d}$. Given $B$ one has the associated
$B$-dephasing map over $L({\cal H})$ given by 
$X\mapsto{\cal D}_{B}(X)=\sum_{i=1}^{d}|i\rangle\langle i|\,\langle i|X|i\rangle$ 
\footnote{One could consider more general dephasing
maps as \unexpanded{${\cal D}(X)=\sum_{i}\Pi_i X\Pi_i$}, where the \unexpanded{$\Pi_{i}$} are
a complete set of projections not necessarily one-dimensional. This
would lead to the theory of coherence with respect to the family of
subspaces \unexpanded{$\mathrm{Im}\Pi_{i}$} \cite{marvian_how_2016}.}. The dephasing map ${\cal D}_{B}$ can be realized
physically as the measurement CP map associated to any non degenerate
observable $H$ diagonal in the basis $B$. 
For any $B,$ the dephasing map is an orthogonal projection over $L({\cal H})$
equipped with the standard Hilbert-Schmidt scalar product $\langle X,\,Y\rangle:=\mathrm{tr}(X^{\dagger}Y)$.
We will denote by ${\cal Q}_{B}:=\1-{\cal D}_{B}$ the complementary
projection of ${\cal D}_{B}$. 
Naturally, one defines $B$-incoherent operators (states) as operators (states)
that are diagonal in the preferred basis $B$.

\textbf{{Definition 1.--}} The set of $B$-incoherent operators
is the range of the $B$-dephasing map, i.e., ${\mathrm{Im}}\,{\cal D}_{B}$.
We will denote the set of $B$-incoherent
states $\rho$ ($\rho\ge0,\,\mathrm{tr}\,\rho=1$) by $I_{B}$.

From the point of view of this definition one can say that ${\cal D}_{B}$
(${\cal Q}_{B}$) projects an operator onto its incoherent (coherent)
component. The set $I_{B}$ is clearly isomorphic to a $(d-1)$-dimensional
simplex spanned by convex combinations of the $|i\rangle\langle i|$, $i=1,\ldots,d$. 
$CP$ maps (all such maps are assumed to be trace preserving
in this paper) ${\cal T}$ mapping $I_{B}$ into itself will play
a distinguished role in this paper. A necessary and sufficient condition
for the invariance of $I_{B}$ under ${\cal T}$ is given by ${\cal T}{\cal D}_{B}={\cal D}_{B}{\cal T}{\cal D}_{B}$
\cite{marvian_how_2016}. However,  
in this paper we will adopt a slightly stronger invariance condition.

\textbf{{Definition 2.--}} A CP map ${\cal T}$ on $L({\cal H})$
will be called $B$-incoherent iff $[{\cal T},\,{\cal D}_{B}]=0$.
We will write ${\cal T}\in CP_{B}$.

Note that $B$-incoherent maps leave
both the subspace of $B$-incoherent operators {\em{and}} its
orthogonal complement ($\cong\mathrm{Ker}\,{\cal D}_{B}=\mathrm{Im}\,{\cal Q}_{B}$)
invariant.
Let us first establish the following, almost obvious, fact.

\textbf{{Proposition 1.--}} A unitary CP map ${\cal U}(X)=UXU^{\dagger}$
(with $U$ unitary)  is $B$-incoherent iff $U|i\rangle=\eta_{i}|\sigma_{U}(i)\rangle$ where
$\sigma_{U}$ is a ($U$-dependent) permutation of $\{1,\ldots,d\}$
and the $\eta_{i}$'s are $U(1)$-phases. $B$-incoherent unitary
maps form a subgroup of $CP_{B}$. 

\begin{figure}[t]
\hspace{-5mm} \includegraphics[scale=0.35]{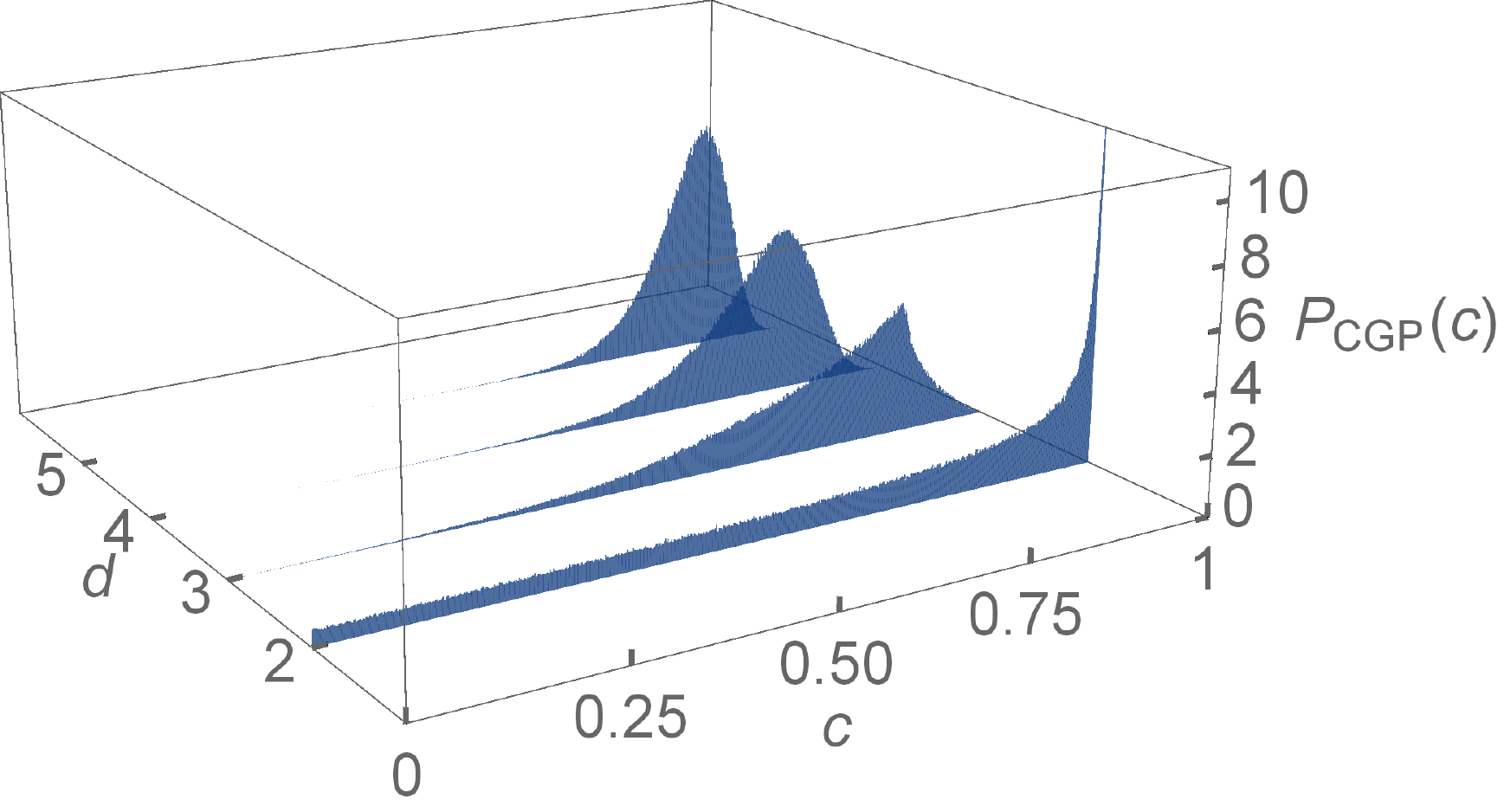} 
\caption{Probability distribution densities (PDD) of the normalized $\tilde{C}_{B}(U)$
for $d=2,\ldots,5$. An ensemble of Haar-distributed $U$'s has
been generated numerically. For $d=2$ the analytical form of the
PDD is $P_{CGP}(c)=\frac{1}{2}(1-c)^{-1/2}$ (see \cite{SM}).}
\label{Fig-PDF-plots} 
\end{figure}

\prlsection{Measures of coherence generating power} Loosely
speaking a coherence measure is a way to quantify how far a given
state is from being incoherent, moreover this quantification is requested
to fulfill some natural properties. More precisely, let us consider
the function $\tilde{c}_{B}(\rho):=\|\rho-{\cal D}_{B}(\rho)\|_{1}=\|{\cal Q}_{B}(\rho)\|_{1}$ ($\Vert X \Vert_1$ denotes the 1-norm of $X$, i.e.~the sum of the singular values of $X$);
this is vanishing iff $\rho$ is $B$-incoherent. Moreover
if ${\cal T}\in CP_{B}$ then $\tilde{c}_{B}({\cal T}(\rho))=\|{\cal Q}_{B}{\cal T}(\rho)\|_{1}=\|{\cal T}{\cal Q}_{B}(\rho)\|_{1}\le\|{\cal Q}_{B}(\rho)\|_{1}=\tilde{c}_{B}(\rho),$
where we have used Definition 2 and the monotonicity of the $1$-norm
under general CP maps. These remarks show that $c_{B}$ is a good
{\em{coherence measure}} with respect to $B$-incoherent operations
\cite{marvian_how_2016}. Unfortunately the $1$-norm is hard to handle
therefore in this paper we will  adopt the Hilbert-Schmidt $2$-norm
$\|X\|_{2}=\sqrt{\langle X,\,X\rangle}$.  We define the 
function 
\begin{equation}
c_{B}(\rho):=\|{\cal Q}_{B}(\rho)\|_{2}^{2}.\label{c_B}
\end{equation}
Again, it is immediate to see that $c_{B}$ vanishes iff $\rho\in I_{B}$
and $\tilde{c}_{B}(\rho)\le\sqrt{d\,{c}_{B}(\rho)}$. On the other
hand it is now not true that $c_{B}$ is necessarily non-increasing
under general $B$-incoherent CP maps (as the $2$-norm does not have
that property either). However if $\cal T$ is unital i.e., ${\cal T}(\1)=\1,$
then $\|{\cal T}(X)\|_{2}\le\|X\|_{2}$ \footnote{One has that 
\unexpanded{$\|{\cal T}(X)\|_2^2=\langle {\cal T}^*{\cal T}(X),\,X\rangle\le \lambda_M \|X\|_2^2$} 
where \unexpanded{$\lambda_M$} is the largest eigenvalue of \unexpanded{$ {\cal T}^*{\cal T}$}. The latter operator is a trace-preserving
CP-map for unital \unexpanded{$\cal T$} and therefore \unexpanded{$\lambda_M\le 1$}. Since the argument of the 2-norm  in Eq.~(\ref{c_B}) is always traceless, for the monotonicy property to hold
suffices that \unexpanded{${\cal P}_0  {\cal T}^*{\cal T} {\cal P}_0$ (${\cal P}_0$}
projection over the space of traceless operators) has eigenvalues smaller than one. 
This property is weaker than unitality.}. Thereby
the desired monotonicity property is recovered if one restricts to
the set of unital $B$-incoherent CP maps. 

Let us now introduce the main novel concept of this paper

\textbf{{Definition 3.--}} The coherence generating
power (CGP) $C_{B}(U)$ of a unitary CP map $X\mapsto{\cal U}(X):=UXU^{\dagger},\,(U\in U({\cal H}))$
with respect the basis $B$ is defined as 
\begin{equation}
C_{B}(U):=\langle c_{B}({\cal U}_{off}(|\psi\rangle\langle\psi|))\rangle_{\psi}\label{C_B}
\end{equation}
where ${\cal U}_{off}:={\cal Q}_{B}{\cal U}{\cal D}_{B}$ and the
average over $\psi$ is taken according to the Haar measure.

The operational idea behind our definition (\ref{C_B}) of CGP is
simple: the power of a unitary $U$ to generate coherence (in a preferred
basis $B$) is given by the average coherence, as measured by the
function (\ref{c_B}), obtained by $U$ acting over an ensemble of
incoherent states. The latter is prepared by a stochastic process
that involves first the generation of (Haar) random quantum states,
and then their $B$-dephasing e.g., by performing a non-selective
measurement of any non-degenerate $B$-diagonal observable. Note that
the ensemble so generated coincides with the uniform one over the
simplex $I_B$ (see \cite{SM}). Of course
other definitions are possible. For example, besides the freedom of
choosing a coherence measure different from (\ref{c_B}), one might
have resorted to a different ensemble of $B$-diagonal states or even
replace the average by a supremum over the ensemble \cite{mani_cohering_2015,bromley_frozen_2015,garcia-diaz_note_2015}. However, our
choice, thanks to the high symmetry of the Haar measure, will allow
us to establish properties of CGP on general grounds as well as to
compute it in an explicit analytic fashion. The most basic properties
of the CGP can be derived directly from Eq.~(\ref{C_B}).

\textbf{{Proposition 2.--}} \textbf{{a)}}
$C_{B}(U)\ge0$ and $C_{B}(U)=0$ iff ${\cal U}\in CP_{B}$. \textbf{{b)}}
If W is a unitary such that ${\cal W}\in CP_{B}$ then $C_{B}(UW)=C_{B}(WU)=C_{B}(U)$.
\textbf{{c)}} Let $\{|\tilde{i}\rangle:=V|i\rangle\}_{i=1}^{d}$
be a new basis $\tilde{B}:=BV$ obtained from $B$ by the (right) action of the unitary $V$ then: 
$C_{BV}(U)=C_{B}(V^{\dagger}UV)$.

Part b) shows that CGP does not change if the ensemble is pre- or
post-processed by incoherent unitaries. Moreover, Part c) shows that
computing the CGP for a single {\em{given}} basis $B_{0}$ is
in principle sufficient for obtaining it for {\em{any}} basis
$B$ (for, given any pair of bases, there is always a unitary connecting
them). It also implies, as we will see, that the statistical properties
of the CGP over the unitary group are {\em{universal}} in the
sense of being basis independent: just the Hilbert space dimension
$d$ matters.

It is important to stress that Prop. 2 holds for a more general choice of $c_B$ than Eq.~(\ref{c_B}) e.g., for $\tilde{c}_B$
\footnote{In fact Prop. 2 holds true by replacing Eq.~(\ref{C_B}) with
\unexpanded{$\tilde{C}_B(U):=\langle D({\cal U}{\cal D}_B(|\psi\rangle\langle\psi|),\,{\cal D}_B{\cal U}{\cal D}_B(|\psi\rangle\langle\psi|))\rangle_\psi$},
where $D$ is any non-negative function over pairs of states such that i) vanishes iff the two arguments are identical, ii) is unitary invariant \unexpanded{$D({\cal U}(\sigma), \,
{\cal U}(\rho))=D(\sigma,\,\rho)$}. For example one could take $D$ to be  the trace distance or the quantum relative entropy.}.  
The choice of the Hilbert-Schmidt norm in the definition of CGP, on the other hand, while
imposing the somewhat severe unitality constraint, has the great advantage of allowing
one for an {\em{explicit}} computation of $C_{B}(U)$. 

\begin{figure}[t]
\begin{centering}
\includegraphics[scale=0.25]{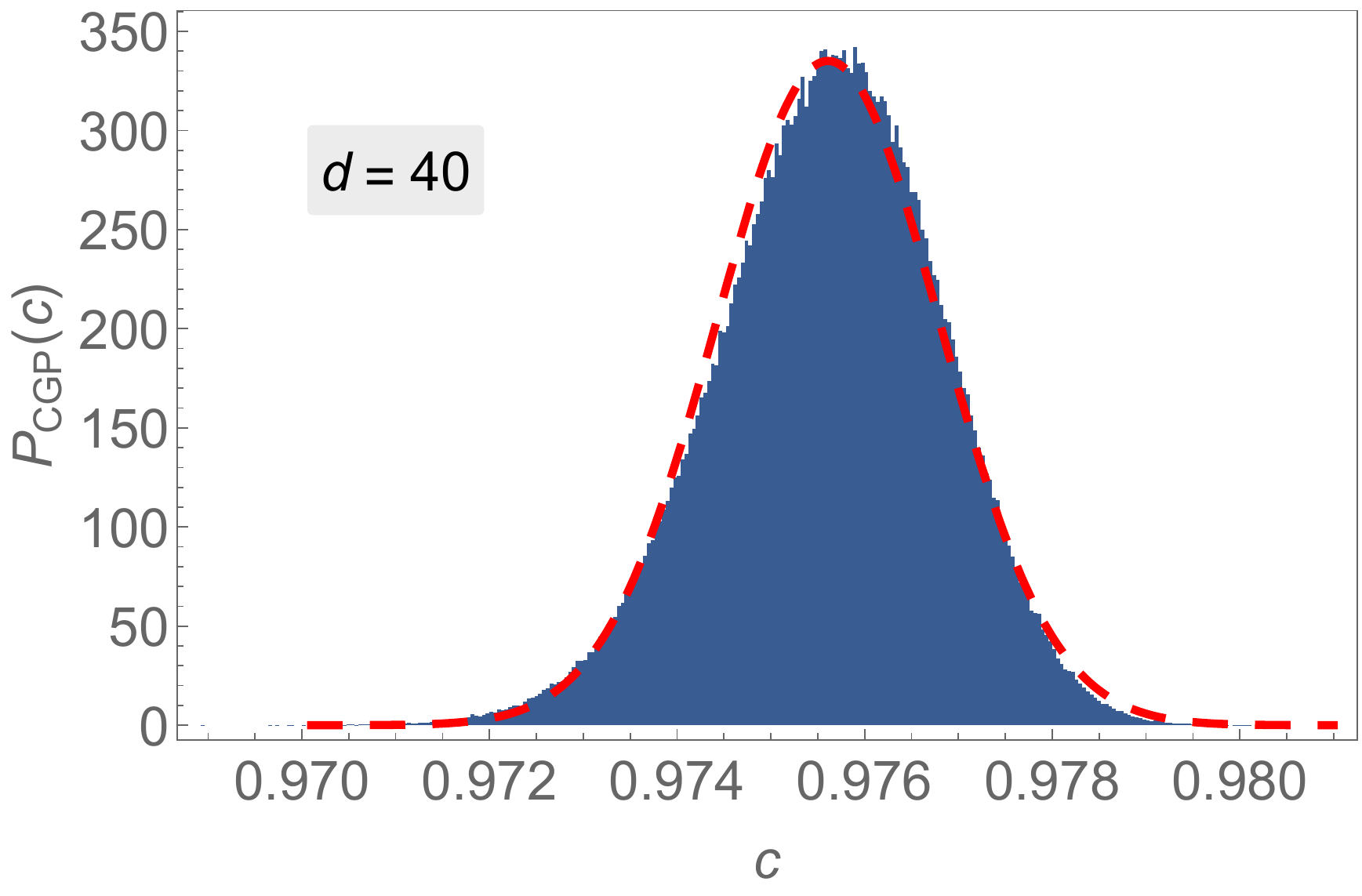} 
\end{centering}
\caption{Probability distribution density (PDD) of the normalized $\tilde{C}_{B}(U)$
for $d=40$. A Gaussian fit is superimposed on the numerically generated
PDD to highlight the central-limit type behavior.}

\label{Gaussian-Fit-d=00003D40} 
\end{figure}


\textbf{{Proposition 3.--}} Let $|\Phi^{+}\rangle=1/\sqrt{d}\sum_{i=1}^{d}|i\rangle^{\otimes\,2}$
be the maximally entangled $d\times d$ singlet, then: \textbf{{a)}}
\begin{equation}
C_{B}(U)=\frac{1}{d+1}\left[1-\mathrm{tr}\,(S\omega_{B}(U))\right],\label{C_B-explicit}
\end{equation}
where $\omega_{B}(U):=({\cal D}_{B}{\cal U}{\cal D}_{B})^{\otimes\,2}(|\Phi^{+}\rangle\langle\Phi^{+}|)$
and $S=\sum_{i,j=1}^{d} |ij\rangle\langle ji|$ is the swap operator over $\mathcal{H}^{\otimes2}$; 
\textbf{{b)}} $\mathrm{tr}\,(S\omega_{B}(U))=1/d\sum_{i,j=1}^{d}|\langle i|U|j\rangle|^{4}$; 
\textbf{{c)}}  $C_{B}(U)\le\frac{1-1/d}{d+1}=:C_{d}$.
The upper bound is saturated iff $|\langle i|U|j\rangle|^{2}=1/d\,(\forall i,j)$.


Part c) of Prop.~3 above shows the  fact that for $U$ to be a unitary
with maximal CGP the base $B$ and the base $BU:=\{U|i\rangle\}_{i=1}^{d}$
have to be {\em{mutually}} unbiased \cite{schwinger_unitary_1960,ivanovic_determination_1993}. For example
the unitary $U$ such that $\langle h|U|m\rangle=1/\sqrt{d}\exp(i\frac{2\pi}{d}hm),\,(h,m=1,\ldots,d)$
has maximal CGP. We also remark that from a) and b) above it follows
easily that $C_{B}(U)=C_{B}(U^{\dagger})$.

Eq.~(\ref{C_B-explicit}) naturally leads to an operational protocol
for the detection of the CGP of a unitary $U$ which does not require
the generation of a Haar distributed ensemble of states or quantum
process tomography \cite{chuang_prescription_1997,poyatos_complete_1997}.

\textbf{{Protocol for CGP detection:}} 
\textbf{{1)}} Prepare $|\Phi^{+}\rangle$; \textbf{{2)}} $B$-dephase both subsystems;
\textbf{{3}}) Apply ${\cal U}$ to both subsystems; \textbf{{4)}}
$B$-dephase again both subsystems; \textbf{{5)}} Measure the expectation
value of the observable $S$; \textbf{{6)}} Plug the obtained value
in Eq.~(\ref{C_B-explicit}). This protocol is depicted in Fig.~(\ref{protocol-Fig}).
Since 
\begin{equation}
{\cal D}_{B}^{\otimes\,2}(|\Phi^{+}\rangle\langle\Phi^{+}|)=\frac{1}{d}\sum_{i=1}^{d}|i\rangle\langle i|^{\otimes\,2}=:\rho_{B},\label{rho_B}
\end{equation}
steps 1) and 2) above can be replaced by 1') Prepare the maximally
classically $B$-correlated state $\rho_{B}$ (for which is enough
to $B$-dephase one subsystem). This shows that entanglement is not
really needed in the detection of $C_{B}(U)$. However, in 5) one
is required to measure $S$ which involves non-trivial interactions
between the two $d$-dimensional subsystems. This is the experimentally
more challenging part of the protocol. Notice, however, that for two-qubits, this amounts to a standard Bell's basis measurement. 
\begin{figure}[!hb]
\includegraphics[scale=0.305]{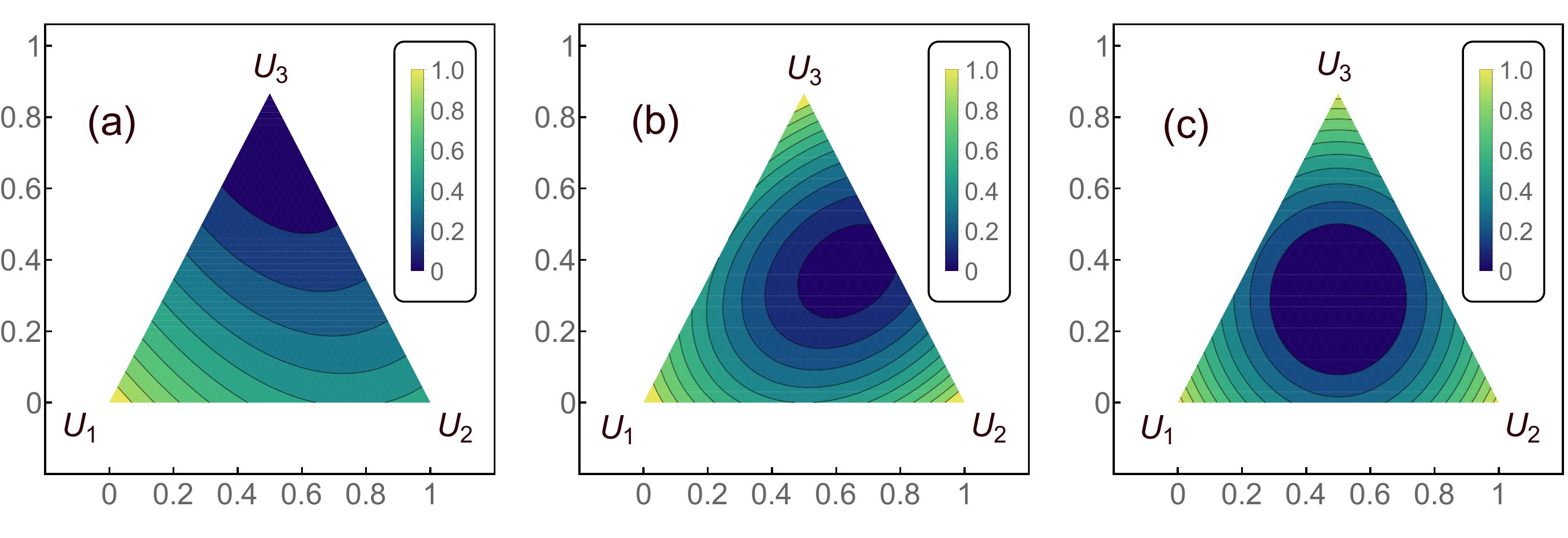} \caption{Coherence generating power for convex combinations of unitaries of the form $\mathcal{E}(\cdot)=\sum_{k=1}^3 p_{k}U_{k} \cdot {U_{k}}^{\dagger}$. One always has the convexity inequality 
$\tilde{C}_B(\mathcal{E}) \le \sum_{k=1}^3 p_{k} \tilde{C}_B(U_k)$ as noted in the main text. 
\textbf{(a)} Here $d=3$ and we fixed $U_k$ such that $\tilde{C}_{B}(U_{1})=1,\tilde{C}_{B}(U_{2})=1/2$ and $\tilde{C}_{B}(U_{3})=0$.
\textbf{(b)} For this example ($d=10$)
$U_{1}$ is the Discrete Fourier Transform matrix  $\langle l | U_{1} |m \rangle = d^{-1/2} \exp{(i lm  2\pi/d)}$ while
$U_{2}$ ($U_{3}$) is obtained by interchanging the first (last)
2 rows of $U_{1}$. All $U_k$ have maximal CGP (simplex vertices)
but the CGP of mixtures can drop
significantly. \textbf{(c)} This is a typical case for randomly chosen
unitaries $U_k$ of large dimension (here $d=40$). One observes that $\tilde{C}_B(U_k)$ is nearly maximal consistently with the concentration phenomenon.}
\label{3-triangles} 
\end{figure}


\prlsection{CGP as a random variable over the unitary group} We
now investigate some of the properties of the CGP of Eq.~(\ref{C_B-explicit})
seen as a random variable over the unitary group $U({\cal H})$ equipped
with the Haar measure $d\mu(U).$ 

\textbf{{Proposition 4.--}} \textbf{{a)}}
The probability distribution density $P_{CGP}(c):=\int_{U({\cal H})}d\mu(U)\,\delta(c-C_{B}(U))$
for the CGP Eq.~(\ref{C_B-explicit}) is independent of $B$. \textbf{{b)}}
The first moment is given by 
\begin{equation}
\langle C_{B}(U)\rangle_{U}=\int dcP_{CGP}(c)=\frac{d-1}{(d+1)^{2}}. \label{C_B-average}
\end{equation}
\textbf{{c)}} Let us define the normalized CGP $\tilde{C}_{B}(U):={C}_{B}(U)/C_{d}\le1$
then $\langle\tilde{C}_{B}(U)\rangle_{U}=(1+1/d)^{-1}$. Using Levy's
lemma for unitaries \cite{anderson_introduction_2009} one obtains 
\begin{equation}
\mathrm{Prob}\left(\tilde{C}_{B}(U)\ge1-2/d^{1/3}\right)\ge1-\exp\left(-d^{1/3}/256\right).\label{Levy}
\end{equation}


Eq.~(\ref{Levy}) shows that in high-dimension a random unitary will
have, with overwhelming probability, nearly maximal CGP. In Fig.~\ref{Fig-PDF-plots} 
are reported numerical simulations of the probability
distribution function of $\tilde{C}_{B}(U)$ for Haar distributed
$U$ in different dimensions. In particular numerics shows that
the variance of $\tilde{C}_{B}(U)$ is $O(1/d^{3})$ (see \cite{SM}). Moreover,
Fig.~\ref{Gaussian-Fit-d=00003D40} shows that for large Hilbert
space dimension $d$ a central-limit type behavior emerges and the
$P_{CGP}$ can be well approximated by a normal distribution.

\prlsection{Beyond Unitarity and finite dimensions} In this section
we will briefly discuss how our approach extends to CP maps that are
not necessarily unitary and how one might extend our formalism to
infinite dimensions e.g., optical modes. Since we still would like
to employ the Hilbert-Schmidt norm we will focus here on {\em{unital}}
maps. We can still adopt Eq.~(\ref{c_B}) for the definition of a
coherence measure. Moreover, incoherent (according to Def.~2) ${\cal E}$
will not increase it. We can now define the CGP of ${\cal E}$ by
the same Eq.~(\ref{C_B}) (with ${\cal E}$ replacing ${\cal U}$).
It is still true that $C_{B}({\cal E})=0\Leftrightarrow{\cal E}_{off}:={\cal Q}_{B}{\cal E}{\cal D}_{B}=0$
but this is now a weaker property than incoherence as it does not imply
$[{\cal E},\,{\cal D}_{B}]=0$. The corresponding measure of CGP is
therefore not {\em{faithful}} i.e., $C_{B}({\cal E})=0\Rightarrow{\cal E}\in CP_{B}$
doesn't hold (just the converse does) \footnote{If one insists that $C_{B}$ should be faithful
then the set of incoherent maps ${\cal E}$ should be defined by ${\cal E}_{off}=0$.
However, in this case the monotonicity property might be violated.
Notice that for normal maps $[{\cal E},\,{\cal E}^{*}]=0$ the two different
notions of incoherence coincide (see part a) in Proof of Prop.~2).}. The following
proposition shows how Prop.~3 generalizes to unital maps more general
than unitaries. 

\textbf{{Proposition 5--}} Let ${\cal E}(\cdot)=\sum_{k}A_{k}\cdot A_{k}^{\dagger},\,(\sum_{k}A_{k}^{\dagger}A_{k}=\1)$
be a unital CP-map over $\mathrm{L}({\cal H})$. If we define its CGP by
Eq.~(\ref{C_B}) (replacing ${\cal U}$ with ${\cal E}$) then it
follows that \textbf{{a)}} $C_{B}({\cal E})\ge0$ and it vanishes
if ${\cal E}$ is $B$-incoherent. \textbf{{b)}} If ${\cal T}$
is $B$-incoherent then $C_{B}({\cal T}{\cal E})\le C_{B}({\cal E})$.
\textbf{{c)}} 
\begin{equation}
C_{B}({\cal E})=\frac{1}{d+1}\left[\mathrm{tr}\,(S\tilde{\omega}_{B}({\cal E}))-\mathrm{tr}\,(S\omega_{B}({\cal E}))\right]\le C_{d},\label{C_B-Texplicit}
\end{equation}
where $\omega_{B}({\cal E}):=({\cal D}_{B}{\cal E})^{\otimes\,2}(\rho_{B})$
and $\tilde{\omega}_{B}({\cal E}):={\cal E}^{\otimes\,2}(\rho_{B})$.
\textbf{{d)}} $C_{B}({\cal E})=[d(d+1)]^{-1}\sum_{i,l\neq m=1}^{d}|\sum_{k}(A_{k})_{li}(A_{k})_{mi}^{*}|^{2}$

Property b) in Prop.~5 is the analog of Eq.~(\ref{C_B-explicit})
and can be similarly interpreted by an operational protocol involving
the measurement of $S$ over the states $\omega_{B}({\cal T})$ and
$\tilde{\omega}_{B}({\cal T})$. Point d) above gives the CGP explicitly
as a function of the matrix elements of the Kraus operators of ${\cal E}$;
it corresponds to b) in Prop.~3. We also note that the function ${\cal E}\mapsto C_{B}({\cal E})$
is {\em{convex}} (since it is a convex combination of the convex
functions ${\cal E}\mapsto c_{B}({\cal E}_{off}(|\psi\rangle\langle\psi|))\,\forall|\psi\rangle$).
It follows that the maximum CGP of a convex set of maps will be achieved
over extremal points. This phenomenon can be seen in the in Fig.~\ref{3-triangles}.

Remarkably, Eq.~(\ref{C_B-Texplicit})
seems to suggest a natural way in which our results can be extended
to infinite dimensions. Let us consider, for simplicity, the unitary
case and normalize Eq.~(\ref{C_B-explicit}) by dividing by $C_{d}$.
Now sending $d\to\infty$ the $d$-dependent pre-factor of CGP disappears
and one is led to consider the expression $\tilde{C}_{B}^{(\infty)}(U)=1-\mathrm{tr}\,(S\omega_{B}^{(\infty)}({U}))$
with $\omega_{B}^{(\infty)}({U})=({\cal D}_{B}{\cal U})^{\otimes\,2}(\rho_{B}^{(\infty)})$
where $\rho_{B}^{(\infty)}$ is {\em{some}} infinite-dimensional
generalization of the maximally classically $B$-correlated state
Eq.~(\ref{rho_B}). For example, for any $\lambda\in(0,\,1),$ one
could choose $\rho_{B}^{(\infty)}:=(1-\lambda^{2})\sum_{i=0}^{\infty}\lambda^{2i}|i\rangle\langle i|^{\otimes\,2}$
\footnote{This is the $B$-dephased of the 2-mode EPR pair \unexpanded{$|\psi_{EPR}\rangle=\sum_{\i=0}^{\infty}\sqrt{1-\lambda^{2}}(-\lambda)^{n}|i\rangle^{\otimes\,2}$}
with squeezing parameter $\tanh^{-1}(\lambda)$.}. With this choice it is immediate to check that $\tilde{C}_{B}^{(\infty)}(U)=0$
iff $U$ is incoherent and that post-processing with incoherent unitaries
leaves the CGP invariant \footnote{Invariance under pre-processing by incoherent
unitaries does not hold because, for $\lambda<1$ it is not true that
$[U^{\otimes\,2},\,\rho_{B}]=0$ for all such $U$'s. One has to consider
the non-trivial limit $\lambda\to1$.}. Developments in the
infinite-dimensional case will be presented elsewhere \cite{future}.


\prlsection{Asymmetry}
Closely related to the theory of coherence is the notion of {\em{asymmetry}}
\cite{marvian_asymmetry_2014,marvian_extending_2014,marvian_mashhad_symmetry_2012}. Given an
observable $H$ 
one says that a state $\rho$ (CP map ${\cal E}$) is $H$-symmetric
($H$-covariant) iff $[H,\,\rho]=:{\cal H}(\rho)=0$ ($[{\cal H},\,{\cal E}]=0$).
An {\em{asymmetry measure}} is a real valued function $a_{H}(\rho)$
that vanishes over symmetric states and is non-increasing under covariant
CP maps i.e., $a_{H}({\cal E}(\rho))\le a_{H}(\rho)$ \cite{marvian_mashhad_symmetry_2012}.
Following the main idea of this paper one could define the asymmetry
generating power (AGP) of a CP map ${\cal E}$ by $A_{H}({\cal E}):=\langle a_{H}({\cal E}(\omega))\rangle_{\omega}$
where the average is performed over a suitable ensemble of $H$-symmetric
states $\omega$. 
First results in this direction and connection between AGP and the CGP defined in this paper are discussed in \cite{SM}.

\prlsection{Conclusions}
In this paper we have discussed a way to quantify the coherence generating
power (CGP) of a quantum operation. As a coherence measure we have
conveniently adopted the Hilbert-Schmidt norm of the coherent part
of a quantum state. Our approach is to look at the average coherence
produced when the operation is performed over a uniform ensemble of
input incoherent states. The input ensemble is obtained by dephasing,
with respect to the chosen basis, an ensemble of pure states distributed
according the Haar measure and coincides with the uniform measure over the simplex of states 
spanned by the pure basis states. 

Under these assumptions one obtains an
analytically computable measure of CGP for arbitrary unital operations in any dimension. Operational protocols for
the direct detection of CGP have been described. Neither the ability
to generate the Haar distributed input ensemble nor quantum process
tomography are required. We  focused on unitary maps, characterized
those with maximal CGP, studied the distribution of this measure over
the unitary group, both analytically and numerically. For unitary
maps this distribution is universal (basis independent) and for large
Hilbert space dimension a central-limit type phenomenon emerges. A
random unitary has, with overwhelming probability, nearly maximal
CGP. Finally, we extended our approach to quantify the power of an
operation to generate a more general type of asymmetry. 

The analytical
framework  here established is particularly suited for unital
quantum maps. Going beyond unitality, finite dimensionality, and extending  to general resource theories
represent challenging tasks for future investigations.

{\em{Acknowledgements.- }} This work was partially supported
by the ARO MURI grants W911NF-11-1-0268 and W911NF-15-1-0582. P.Z. thanks I.~Marvian for
introducing him to asymmetry measures and F.G.S.~L.~Brand\~{a}o
for pointing out the right form of the Levy's Lemma. 

%



\appendix

\section{Proof of Proposition 1}

This condition is clearly a sufficient for
$B$-incoherence as commutativity of ${\cal U}$ and ${\cal D}_{B}$
can be explicitly checked in a straightforward fashion. It is also
necessary. Indeed if ${\cal U}$ is $B$-incoherent then ${\cal U}(|i\rangle\langle i|)$
must be $B$-diagonal for all $i;$ because of unitarity, it also
must be a one-dimensional projector whence ${\cal U}(|i\rangle\langle i|)$
has necessarily the form $|j\rangle\langle j|$ where $|j\rangle$
is a uniquely defined element $j=:\sigma_{U}(i)$ of $B$. The only
degrees of freedom of $U$ left are then $U(1)$-phases. The last
statement of the proposition is evident. $\hfill\Box$

\section{Equivalence of ensembles}

Here we show that the ensemble constructed in the main text coincides
in fact with the uniform distribution over the simplex spanned by
the states $|i\rangle\langle i|$, $i=1,\ldots, d$. For any (measurable) function $f$
the expectation value over the ensemble is given by $\langle f(\mathcal{D}_{B}(|\psi\rangle\langle\psi|)\rangle_{\psi}$.
Calling $\psi_{i}=\langle i|\psi\rangle$ and $p_{i}=|\langle i|\psi\rangle|^{2}$ we can write it as 
\begin{multline}
\langle f(\mathcal{D}_{B}(|\psi\rangle\langle\psi|)\rangle_{\psi}=M\int d\psi_{1}\cdots\int d\psi_{d}\, \times \\
\times f(p_{1},\ldots,p_{d})\delta(1-\sum_{i=1}^{d}p_{i})
\end{multline}
where $M$ is a normalization constant and $d\psi_{i}=d\mathrm{Re}(\psi_{i})\,d\mathrm{Im}(\psi_{i})$.
Switching to polar coordinates one has $d\psi_{i}=r_{i}dr_{i}\,d\vartheta_{i}=dp_{i}\,d\vartheta_{i}/2$.
Performing the integration over the angles $\vartheta_{i}$ we obtain
\begin{multline}
\langle f(\mathcal{D}_{B}(|\psi\rangle\langle\psi|)\rangle_{\psi}=M'\int dp_{1}\cdots\int dp_{d}\, \times \\
\times f(p_{1},\ldots,p_{d})\delta(1-\sum_{i=1}^{d}p_{i})\,,
\end{multline}
that is, the uniform measure over the simplex ($M'$ is another normalization
constant).

\section{Proof of Proposition 2}

a) By definition the CGP is non-negative, moreover
$C_{B}(U)=0$ implies ${\cal U}_{off}(|\psi\rangle\langle\psi|)=0,\forall|\psi\rangle,$
which in turn implies that ${\cal U}_{off}={\cal U}{\cal D}_{B}-{\cal D}_{B}{\cal U}{\cal D}_{B}=0$.
This equation, as remarked in the above, shows that $\mathrm{Im}\,{\cal D}_{B}$
is invariant under ${\cal U}$ but, since ${\cal U}$ is normal, also
the orthogonal complement $\mathrm{Ker}\,{\cal D}_{B}$ is invariant.
It follows that $[{\cal U}_{B},\,{\cal D}_{B}]=0$ that is what we
wanted to prove. b) $C_{B}(WU)=C_{B}(U)$ ($C_{B}(UW)=C_{B}(U)$)
follows from the commutativity of ${\cal W}$ and ${\cal Q}_{B}$
(${\cal D}_{B}$) and the unitary invariance of the Hilbert-Schmidt
norm (Haar measure). c) By definition of $\tilde{B}=BV$ one finds
${\cal D}_{\tilde{B}}={\cal V}{\cal D}_{B}{\cal V}^{\dagger},$ ($\mathcal{V} (\cdot) = V \cdot V^\dagger$) inserting
this relation in Eq.~(2) in the main text and using again unitary invariance
of the Hilbert-Schmidt norm and of the Haar measure one completes
the proof. $\hfill\Box$

\section{Lemma}

If $S$ is the swap operator over ${\cal H}^{\otimes\,2}$ ($S=\sum_{i,j=1}^{d}|ij\rangle\langle ji|,\,\,{\cal H}=\mathrm{span}\{|i\rangle\}_{i=1}^{d}$)
then: a) $\|X\|_{2}^{2}=\mathrm{tr}\,\left(SX\otimes X\right);$ b)
$\langle|\psi\rangle\langle\psi|^{\otimes\,2}\rangle_{\psi}=[d(d+1)]^{-1}(\1+S)$ where the average is taken over Haar distributed $\psi$ in $\mathcal{H}$,
(see e.g.,~\cite{keyl_review}).

\section{Proof of Proposition 3}

a) Using the {\em{Lemma}} and Definition
3 one can immediately write $C_{B}(U)=[d(d+1)]^{-1}\mathrm{tr}\left[S\,{\cal U}_{off}^{\otimes\,2}(\1+S)\right]$.
The first term in this expression is vanishing; indeed ${\cal U}_{off}^{\otimes\,2}(\1)={\cal Q}_{B}^{\otimes\,2}(\1)=0$
(the identity is a diagonal operator for any $B$). Using the fact that
$\forall Y$ one has $\mathrm{tr}\left[S{\cal Q}_{B}^{\otimes\,2}Y\right]=\mathrm{tr}\left[S(\1-{\cal D}_{B}^{\otimes\,2})Y\right]$
the second term can be written as $\mathrm{tr}\left[S\,{\cal U}_{off}^{\otimes\,2}(S)\right]=\mathrm{tr}\left[S\,(\1-{\cal D}_{B}^{\otimes\,2})({\cal U}{\cal D}_{B})^{\otimes\,2}(S)\right]$
Moreover, ${\cal D}_{B}^{\otimes\,2}(S)=\sum_{i=1}^{d}|i\rangle\langle i|^{\otimes\,2}=:d\rho_{B}$
therefore the first term in the last equation can be now written as
$(d+1)^{-1}\mathrm{tr}\left(S{\cal U}^{\otimes\,2}(\rho_{B}))\right)=(d+1)^{-1}$.
The last equality follows from the fact that ${\cal U}^{\otimes\,2}(\rho_{B})$
is entirely supported in the eigenvalue one subspace of $S$ (symmetric
subspace). Observing now that is also true that ${\cal D}_{B}^{\otimes\,2}(S)=d{\cal D}_{B}^{\otimes\,2}(|\Phi^{+}\rangle\langle\Phi^{+}|)$
completes the proof of part a). Let us now move to part b). One has
$\mathrm{tr}\left(S({\cal D}_{B}{\cal U}^{\otimes\,2})(\rho_{B})\right)=1/d\sum_{i=1}^{d}\mathrm{tr}\left(S({\cal D}_{B}{\cal U})^{\otimes\,2}(|i\rangle\langle i|^{\otimes\,2})\right)=1/d\sum_{i=1}^{d}\|({\cal D}_{B}{\cal U}(|i\rangle\langle i|)\|_{2}^{2}$.
But ${\cal D}_{B}{\cal U}(|i\rangle\langle i|)=\sum_{j=1}^{d}|\langle i|U|j\rangle|^{2}|j\rangle\langle j|$.
Bringing together the last two equations completes the Proof of part b).
Now part c). From the above one sees that $d\mathrm{tr}\,(S\omega_{B}(U))$
is the sum of $d$ purities $\|{\cal D}_{B}{\cal U}(|i\rangle\langle i|)\|_{2}^{2},\ (i=1,\ldots,d)$.
Therefore the minimum of this quantity occurs when they are all their minimum
i.e., $1/d$. Adding over $i$ one finds $\langle S\rangle_{\omega_{B}(U)}\ge1/d$
from which the desired upper bound c) follows, This bound is achieved
iff ${\cal D}_{B}{\cal U}(|i\rangle\langle i|)=\1/d,\,(\forall i)$.
This, in turn, from the expression a few lines above, implies $|\langle i|U|j\rangle|^{2}=1/d$.
Notice that this conclusion can be also derived directly from the
formula c). $\hfill\Box$

\section{Proof of Proposition 4}

a) Given a fixed basis $B_{0}$ and any other
base $B$ one has that there exists a $V\in U({\cal H})$ such that
$B=B_{0}V$ (see comment after Prop.~2 in the main text). Therefore
$P_{B}(c)=P_{B_{0}V}(c)dc=\int d\mu(U)\,\delta(c-C_{B_{0}V}(U))=\int d\mu(U)\,\delta(c-C_{B_{0}}(V^{\dagger}UV))=\int d\mu(U)\,\delta(c-C_{B_{0}}(V^{\dagger}UV))=\int d\mu(VWV^{\dagger})\,\delta(c-C_{B_{0}}(W))=P_{B_{0}}(c)dc$.
Where we have used c) of Prop.~2 and the unitary invariance of the
Haar measure i.e., $d\mu(VWV^{\dagger})=d\mu(V)$. b) Let us consider
the terms $|\langle i|U|j\rangle|^{4}$ from part b) of Prop. 3 and
perform average with respect a Haar distributed $U$. Denoting by
$|\psi\rangle=U|j\rangle$ this amounts to average with respect $|\psi\rangle$
the following quantity $(\langle i|\psi\rangle\langle\psi|i\rangle)^{2}=\mathrm{tr}\,\left(|i\rangle\langle i|^{\otimes\,2}|\psi\rangle\langle\psi|^{\otimes\,2}\right)$.
Using now the Lemma one finds $\langle|\langle i|U|j\rangle|^{4}\rangle_{U}=\langle|\langle i|\psi\rangle|^{4}\rangle_{\psi}=[d(d+1)]^{-1}\mathrm{tr}\,\left(|i\rangle\langle i|^{\otimes\,2}(\1+S)\right)=2[d(d+1)]^{-1}$.
Adding over $i$ and $j$ and using Eq.~(3) in the main text one
obtains Eq.~(5). c) Here we need a version of the
Levy Lemma formulated for Haar distributed $d\times d$ unitaries:
Prob$\{X(U)-\langle X(U)\rangle_{U}\ge\epsilon\}\le\exp\left[-\frac{d\epsilon^{2}}{4K^{2}}\right]$
where $K$ is a Lipschitz constant of $X\colon U(d)\mapsto{\mathbf{R}}$
i.e., $|X(U)-X(V)|\le K\|U-V\|_{2}$ \cite{haar_U}. Let us set $X(U):=1-\tilde{C}_{B}(U)$
then $X(U)-\langle X(U)\rangle_{U}=1-\tilde{C}_{B}(U)-1/(d+1)$ from
which Prob$\{\tilde{C}_{B}(U)\le1-\epsilon-1/d\}\le\exp(-d\epsilon^{2}/(4K^{2}))$.
If we now set $\epsilon=d^{-\alpha}$ with $\alpha\in(0,\,1/2)$ we
get 
\[
\mathrm{Prob}\{\tilde{C}_{B}(U)\le1-2/d^{\alpha}\}\le\exp(-d^{1-2\alpha}/(4K^{2})).
\]
To complete the proof we have to estimate the Lipschitz constant $K$.
For this, from Eq.~(3), and the definitions above,
is clearly enough to consider the function $f(U)=1/d\sum_{i=1}^{d}{\mathrm{tr}}\left(S({\cal D}_{B}^{\otimes\,2}(|i_{U}\rangle\langle i_{U}|^{\otimes\,2})\right)=:1-d/(d-1)\tilde{C}_{B}(U)$
where $|i_{U}\rangle:=U|i\rangle$. Let us consider each of the $d$
terms, called $f_{i}(U),$ separately: $|f_{i}(U)-f_{i}(V)|\le|{\mathrm{tr}}\left(S{\cal D}_{B}^{\otimes\,2}(|i_{U}\rangle\langle i_{U}|^{\otimes\,2}-|i_{V}\rangle\langle i_{V}|^{\otimes\,2})\right)|\le\||i_{U}\rangle\langle i_{U}|^{\otimes\,2}-|i_{V}\rangle\langle i_{V}|^{\otimes\,2}\|_{1},$
where we have used $\mathrm{tr}(AB)\le\|A\|_{\infty}\|B\|_{1},$ $\|S\|_{\infty}=1$
and, since $B$-dephasing is a CP map, $\|{\cal D}_{B}^{\otimes\,2}(X)\|_{1}\le\|X\|_{1}$.
Now, the last trace-norm distance can be upper bounded by twice the
Hilbert space distance $\||i_{U}\rangle^{\otimes\,2}-|i_{V}\rangle^{\otimes\,2}\|\le\|U^{\otimes\,2}-V^{\otimes\,2}\|_{\infty}=\|1-(U^{\dagger}V)^{\otimes\,2}\|_{\infty}$.
if $\Delta:=U-V$ and $K:=U^{\dagger}\Delta$ has $U^{\dagger}V=1-K$
and then the last norm becomes $\|1-(1-K)^{\otimes\,2}\|_{\infty}=\|K\otimes\1+\1\otimes K+K\otimes K\|_{\infty}\le\|K\|_{\infty}(2+\|K\|_{\infty})\le4\|K\|\le4\|U-V\|_{\infty}\le4\|U-V\|_{2}$
where we have used standard operator norm inequalities. Bringing all
together $\|f(U)-f(V)\|\le8\|U-V\|_{2}$ showing that one can take
$K=8$. Setting $\alpha=1/3$ and considering he complementary inequality
one obtains Eq.~(6) in the main text. $\hfill\Box$ 

\section{Scaling of the Variance}
\begin{figure}[h]
\hspace{-5mm} \includegraphics[scale=0.4]{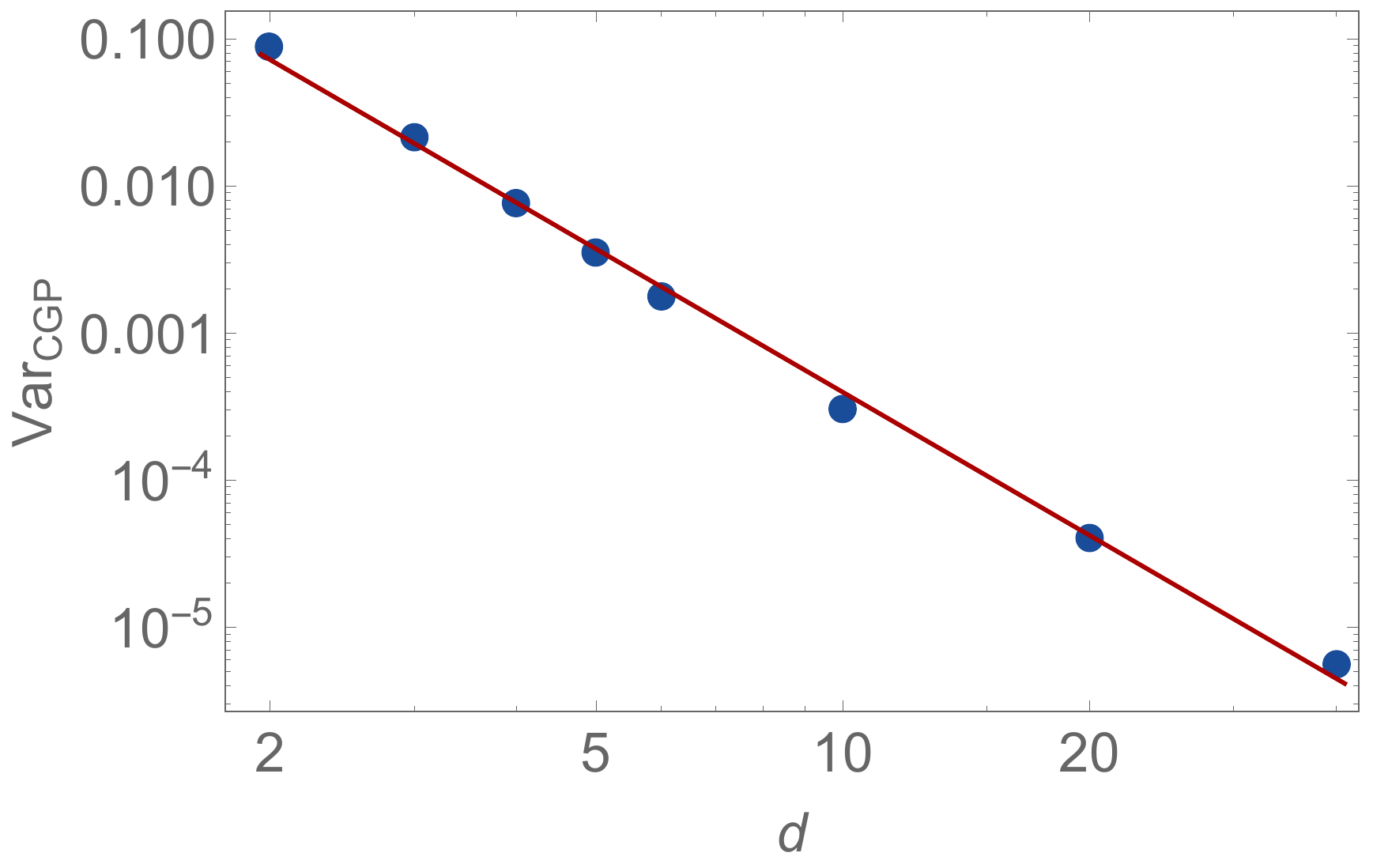} 
\caption{
This log-log plot shows the numerically computed
variance of the random variable $\tilde{C}_{B}(U)$ (where $U$ is
distributed according to the Haar measure) for different values of
the dimension $d$ of the Hilbert space. A power-law $A/d^{\alpha}$
least square fitting (taking into account only the points $d=6,10,20,40$)
gives $\alpha=3.01$, suggesting that the variance of $\tilde{C}_{B}(U)$
is $O(1/d^{3})$.
}
\label{fig-scaling}
\end{figure}

See Fig.~(\ref{fig-scaling}).

\section{Proof of Proposition 5}

Proceed exactly as in the unitary case. The only
difference is that, for general ${\cal E}$ the state ${\cal E}^{\otimes\,2}(\rho_{B})=\tilde{\omega}_{B}$
is not entirely supported in the eigenvalue one eigenspace of $S$.$\hfill\Box$

\section{PDD for CGP in $d=2$}

Using Eq.~(3) for a $SU(2)$
matrix one finds $C_{B}(U)=\frac{1}{3}(1-|a|^{4}-|b|^{4}),$ where
$a=\langle0|U|0\rangle=\langle1|U|1\rangle^{*},\,b=\langle1|U|0\rangle=-\langle0|U|1\rangle^{*}$.
Since $|a|^{2}+|b|^{2}=1$ one can use the Bloch sphere parametrization
$|a|=\cos(\theta/2),\,|b|=\sin(\theta/2)$ from which it follows $\tilde{C}_{B}(U)=C_{B}(U)/C_{d=2}=\sin^{2}(\theta)$.
The distribution density of $c=\tilde{C}_{B}(U)\in[0,\,1]$ is given
by 
\begin{eqnarray*}
P_{CGP}(c) & = & \frac{1}{4\pi}\int_{0}^{2\pi}d\phi\int_{0}^{\pi}d(\cos\theta)\delta(c-\sin^{2}\theta)\\
 & = & \frac{1}{2}\int_{-1}^{1}dx\,\delta(c-1+x^{2})=\frac{1}{2\sqrt{1-c}},
\end{eqnarray*}
where we have used $\int dx\,\delta(f(x))=\sum_{x_{0}\,:\,f(x_{0})=0}|f^{\prime}(x_{0})|^{-1}$.

\section{Asymmetry}
In order to directly connect Asymmetry Generating Power (AGP) and and CGP we assume from here on that the Hamiltonian $H$
is {\em{non degenerate}} and that $B=\{|i\rangle\}_{i=1}^{d}$
is the associated basis of eigenvectors. In this case the notion of
$H$-symmetric state and the one of $B$-incoherent collapse. It is
indeed immediate to see that ${\cal H}(\rho)=:[H,\,\rho]=0\Leftrightarrow \mathcal{D}_{B}(\rho)=\rho$
(in the degenerate case incoherence implies symmetry). At the CP map
level, however, one has just that $H$-covariance implies $B$-incoherence
but not the converse. For example unitaries in Prop.~1 realizing
a non-trivial permutation of $B$ are incoherent but not covariant.
As a consequence the set of coherence measures is smaller than the
set of asymmetry measures \cite{Iman-rob-me-pra-2016,Iman-ArXiv-2016}.
We introduce the following notion of AGP for unital maps ${\cal E}$
\begin{equation}
A_{H}({\cal E})=\langle\|{\cal H}{\cal E}{\cal D}_{B}(|\psi\rangle\langle\psi|)\|_{2}^{2}\rangle_{\psi}\label{AGP}
\end{equation}
where once again the average is taken with respect to the Haar measure.
As the CGP Eq.~(7) in the main text (see comment after Prop.~5)
also the AGP Eq.~(\ref{AGP}) is a convex function of its argument. Furthermore,
if $H=\sum_{i=1}^{d}\epsilon_{i}\,|i\rangle\langle i|,\,\delta(H):=\min_{l\neq m}|\epsilon_{l}-\epsilon_{m}|>0$
(non-degeneracy) and $\|{\cal H}\|:=\max_{l\neq m}|\epsilon_{l}-\epsilon_{m}|,$
then the AGP (\ref{AGP}) fulfills the following properties:

\textbf{{Proposition 6.--}} \textbf{{a)}}
$A_{H}({\cal E})=0$ for all incoherent maps ${\cal E}$. In particular
all $H$-covariant maps have vanishing AGP. \textbf{{b)}} if ${\cal T}$
is a unital $H$-covariant map $A_{H}({\cal T}{\cal E})\le A_{H}({\cal E})$.
For unitary $H$-covariant ${\cal T}$ the inequality becomes an equality.
\textbf{{c)}} 
$A_{H}({\cal E})=[d(d+1)]^{-1}\sum_{i,l\neq m}(\epsilon_{l}-\epsilon_{m})^{2}|\langle l|{\cal E}(|i\rangle\langle i|)|m\rangle|^{2}$.
\textbf{{d)}} $\delta^{2}(H)\,C_{B}({\cal E})\le A_{H}({\cal E})\le\|{\cal H}\|^{2}C_{B}({\cal E})$.
\textbf{{e)}} If ${\cal E}(\cdot)=U\cdot U^{\dagger}$
and the unitary $U$'s are Haar distributed then the induced distribution
of $A_{H}(U)$ depends just on the gap spectrum $\{\epsilon_{l}-\epsilon_{m}\}_{l\neq m}.$

{\em{Proof.--}}
a) Follows from ${\cal H}{\cal D}_{B}=0$ and
${\cal E}{\cal D}_{B}={\cal D}_{B}{\cal E}{\cal D}_{B}$ which holds
for incoherent maps. b) Use $[{\cal T},\,{\cal D}_{B}]=0$ for $H$-covariant
maps and the non-increasing property of the Hilbert-Schmidt norm under
unital maps. c) Following the same steps in the proof of a) in Prop.~3
one arrives at $A_{H}({\cal E})=[d(d+1)]^{-1}\sum_{i=1}^{d}\|{\cal H}{\cal E}(|i\rangle\langle i|)\|_{2}^{2}$.
Expanding the norms in this equation and using ${\cal H}(|l\rangle\langle m|)=(\epsilon_{l}-\epsilon_{m})|l\rangle\langle m|)$
one completes the proof. d) From c) using $\delta(H)\le|\epsilon_{l}-\epsilon_{m}|\le\|{\cal H}\|,(\forall l,m)$.
e) If the Hamiltonian eigenbasis is changed by $|i\rangle\mapsto W|i\rangle$
($W$ unitary) then from the result in c) one sees that ${\cal E}\mapsto{\cal W}^{\dagger}{\cal E}{\cal W}$ ($\mathcal{W}(\cdot) = W \cdot W^\dagger$). 
If ${\cal E}(\cdot)=U\cdot U^{\dagger}$ the last equation implies
$U\mapsto W^{\dagger}UW$. The proof can be now completed following
the same reasoning of point c) in the proof of Prop.~2 and observing
that $H$ enters now, having modded the basis away, just through the
differences $\epsilon_{l}-\epsilon_{m}\,(l\neq m=1,\ldots d)$. $\hfill\Box$


\end{document}